\documentclass[a4paper,english]{IEEEtran}
\usepackage[T1]{fontenc}
\usepackage[latin9]{inputenc}
\usepackage{array}
\usepackage{verbatim}
\usepackage{booktabs}
\usepackage{graphicx}

\makeatletter

\pdfpageheight\paperheight
\pdfpagewidth\paperwidth

\providecommand{\tabularnewline}{\\}

\makeatother

\usepackage{babel}
\begin{document}
\title{Large-Scale Classification of Shortwave Communication Signals with
Machine Learning}
\author{Stefan Scholl\\
research@panoradio-sdr.de}
\maketitle
\begin{abstract}
This paper presents a deep learning approach to the classification
of 160 shortwave radio signals. It addresses the typical challenges
of the shortwave spectrum, which are the large number of different
signal types, the presence of various analog modulations and ionospheric
propagation. As a classifier a deep convolutional neural network is
used, that is trained to recognize 160 typical shortwave signal classes.
The approach is blind and therefore does not require preknowledge
or special preprocessing of the signal and no manual design of discriminative
features for each signal class. The network is trained on a large
number of synthetically generated signals and high quality recordings.
Finally, the network is evaluated on real-world radio signals obtained
from globally deployed receiver hardware and achieves up to 90\% accuracy
for an observation time of only 1 second.
\end{abstract}

\section{Introduction}

\subsection{RF Signal Classification}

Automatic radio frequency (RF) signal classification is the task of
identifying the type of an unknown radio signal in the electromagnetic
spectrum (Figure 1). The type or class of a signal is sometimes also
referred to as \emph{mode} or \emph{waveform} and may be defined in
official communication standards (e.g. ITU, Stanag, ICAO), informal
documents or as closed proprietary communication schemes. Signal classification
is mainly used for spectrum monitoring and surveillance applications,
as well as support for cognitive radio operation and dynamic spectrum
access.

The task of signal classification is related to the widely investigated
topic of automatic modulation classification (AMC) \cite{Zhu.2015}.
However, AMC extracts only the generic modulation scheme (e.g. BPSK,
QPSK, 16-QAM, FSK), which is not sufficient to identify the signal
type. Full signal classification requires the consideration of additional
characteristic parameters such as baud rate, symbol shaping, frame
structure or signal envelope. Furthermore, the number of modulation
classes in AMC is often comparably small, because the number of generic
modulation types is rather limited.

In general, classification algorithms can follow two basic approaches:
\begin{itemize}
\item Feature-based: Here, characteristic signal features are manually designed
for each signal class. The features can include statistical properties
of amplitude, instantaneous phase, frequency or other signal parameters.
The classification algorithm can e.g. be based on a set of decision
rules, either manually designed or leant by a classical machine learning
model, such as a decision tree.
\item Deep learning: The classifier model is trained on a large amount of
example data. Features are automatically extracted during the training
process. The classifier is often a deep neural network.
\end{itemize}
RF communications is present at very different frequency bands. Each
band exhibits specific properties, such as the achievable range and
coverage or the available bandwidth. This results in different communication
applications and users and consequently different signal classes in
each band (e.g. satellite at SHF, mobile communications at UHF and
local broadcasting at VHF frequencies). It is therefore useful to
consider the frequency range of interest when designing a signal classifier.
This paper focuses on the shortwave band.

\begin{figure}[h]
\begin{centering}
\includegraphics[width=1\columnwidth]{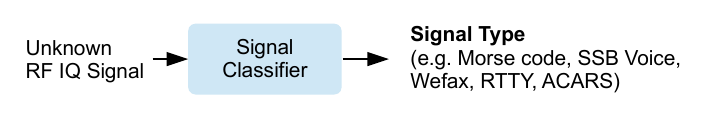}
\par\end{centering}
\centering{}\caption{\label{fig:AMC-vs-classification}Signal classification automatically
determines the type or mode of a received signal.}
\end{figure}

\subsection{Challenges for Shortwave Signal Classification}

The shortwave or high frequency (HF) band covers the radio frequency
range from 3 to 30 MHz and has several advantages over other frequency
bands: It provides potential worldwide coverage even for low transmit
power due to the ionospheric propagation, which reflects radio waves
in the atmosphere \cite{Maslin.1988,Davies.2004}. In addition, shortwave
communication links are independent of large-scale infrastructure
such as satellites, sea cables or relay stations, and can be established
with low-cost transceiver equipment. Due to these advantages, many
operators use this frequency band for long-range communications, including
broadcasting, weather services, aviation, shipping, military, security,
embassies and amateur radio. 

\begin{figure}
\begin{centering}
\includegraphics[width=1\columnwidth]{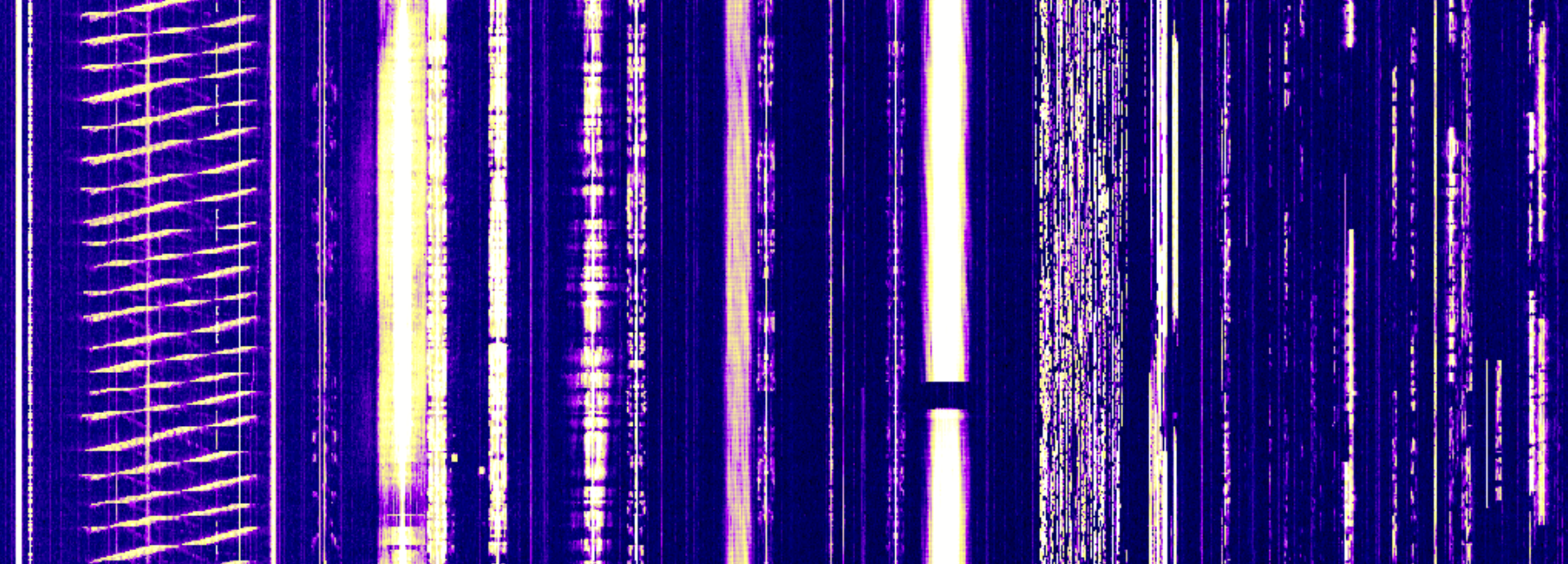}
\par\end{centering}
\caption{\label{fig:Typical-shortwave-spectrum}Exemplary part (here some 100
kHz) of the shortwave spectrum with a variety of different signal
types densely packed and only loosely regulated.}

\end{figure}

Radio signals present in the shortwave band have some differences
from signals in other bands, such as VHF and UHF:
\begin{itemize}
\item \emph{Number of signals:} A large number of different signals from
all over the world can be present in a narrow range of the spectrum
due to the long range coverage, as shown in Figure \ref{fig:Typical-shortwave-spectrum}.
In addition, frequency regulations are comparably loose and difficult
to enforce, resulting in a less well organized spectrum.
\item \emph{Channel models: }The ionospheric propagation of the radio waves
is characterized by a typical time and frequency fading with Doppler
shifts. The exact channel properties are not constant and can vary
over periods of minutes to years. The received noise is often characterized
by atmospheric noise or man-made noise in urban areas \cite{ITU.082024}.
\item \emph{Modulation formats:} Analog modulations types are still widespread,
e.g. single-sideband (SSB) voice, Morse code, HF fax or AM broadcasting.
For digital transmission, M-FSK or modern OFDM modulations are often
used. Higher-order QAM modulations (> 4-QAM) or analog frequency modulation
are rarely seen on shortwave. 
\item \emph{Bandwidth:} Shortwave signals typically cover only a small bandwidth,
often below 4 kHz down to a few Hz.
\end{itemize}
These special properties of shortwave signals present some challenges
for signal classification:
\begin{itemize}
\item A classifier must be able to handle a large number of signals, including
classes with high similarity, for which manual feature design may
be difficult.
\item The special properties of the ionospheric communication channel and
types of noise present must be taken into account. 
\item The recognition of various analog modes must be ensured. This can
be challenging because analog signals often lack clear characteristics
such as a baud rate or well-defined bandwidth, and have more variable
waveform shapes than digital signals.
\end{itemize}

\subsection{Related Work}

Traditional approaches to radio signal classification and AMC rely
on signal features based on probabilistic methods, statistics or cyclostationarity
\cite{Zhu.2015}. These features need to be manually designed by algorithm
developers, which may be costly and difficult for a large number of
different signal classes. Recently, deep learning techniques gained
huge interest and showed good performance for radio signal classification
\cite{Scholl.11.06.2019} and AMC \cite{OShea.2016,TimothyJ.OShea.2017,Boegner.2022}.
These modern machine learning algorithms automatically extract characteristic
features during the training process from labeled data and often do
not require manual feature design. Furthermore, it is possible to
combine deep learning with manually designed features as presented
in \cite{Snoap.2023} for AMC.

For the classification of shortwave signals, several approaches have
been shown in the literature. Feature-based approaches in \cite{Giesbrecht.2016}
and \cite{Dearlove.1999} exploit, for example, statistical and spectral
properties for a small set of five military shortwave waveforms in
\cite{Dearlove.1999} and a set of five miscellaneous FSK, PSK and
AM modes in \cite{Giesbrecht.2016}. Deep learning techniques have
first been applied to shortwave signal classification in \cite{Scholl.11.06.2019}
using convolutional neural networks (CNN) and a Resnet operating on
IQ data to distinguish between 18 typical HF classes (e.g. SSB voice,
different RTTY, Sitor-B, AM, amateur radio modes). A framework for
detecting five different 3G-ALE waveforms using the bi-spectrum and
a CNN has been presented in \cite{Li.2022}. Classification based
on spectrograms and a residual CNN has been shown in \cite{Lin.2024}
for a set of 17 modes (e.g. Clover 2000, Link 11, MS-110A, Pactor).
A regression approach to classification has been introduced in \cite{Zhang.2022}
in order to distinguish between six HF signals (e.g. MS-110A, 2G-ALE,
3G-ALE, Link-11, Pactor). Finally, \cite{Kay.2024} presented an approach
based on permutation entropy for the set of 18 modes from \cite{Scholl.11.06.2019}.
An overview of the current literature on shortwave signal recognition
is provided in Table \ref{tab:Related-Work-sota}.

Although classifiers for shortwave signals face a large number of
different signal classes in real-world operation, only small sets
not exceeding 18 classes have been considered in the literature. However,
a good classifier for the HF band should support a much larger set
of classes, which naturally makes the classification task more challenging.

\begin{table}
\begin{centering}
\begin{tabular}{>{\raggedright}p{1.8cm}>{\raggedright}p{0.5cm}>{\raggedright}p{2.1cm}l>{\raggedright}p{0.8cm}}
\toprule 
Publication & Year & Features & Classifier & Signals Classes\tabularnewline
\midrule
Dearlove \cite{Dearlove.1999} & 1999 & IQ, spectrum & Correlator & 5\tabularnewline
Giesbrecht \cite{Giesbrecht.2016} & 2016 & statistical, spectral & Decision Tree & 5\tabularnewline
Scholl \cite{Scholl.11.06.2019} & 2019 & IQ & CNN, Resnet & 18\tabularnewline
Zhang \cite{Zhang.2022} & 2022 & constellation, bits & CNN & 6\tabularnewline
Li \cite{Li.2022} & 2022 & bi-spectrum & CNN & 5\tabularnewline
Kay \cite{Kay.2024} & 2024 & permut. entropy & CNN & 18\tabularnewline
Lin \cite{Lin.2024} & 2024 & spectrogram & Resnet & 17\tabularnewline
\midrule 
\textbf{This Work} & \textbf{2025} & \textbf{IQ} & \textbf{CNN} & \textbf{160}\tabularnewline
\bottomrule
\end{tabular}
\par\end{centering}
\medskip{}

\caption{\label{tab:Related-Work-sota}Related Work for Shortwave Signal Classification}
\end{table}

\section{Large-Scale Signal Classification with Deep Learning}

\subsection{This Work}

This work presents a signal classifier, that can recognize 160 typical
shortwave modes. For this large number of signal classes, manual feature
design is difficult. Therefore, this work investigates the deep learning
approach and uses a convolutional neural network trained on a large
amount of example data. The approach is blind and does not require
any preknowledge from the signal apart from the training signals themselves.
For a final meaningful evaluation of the classifier, additional real-world
test data is used, that has been recorded from deployed receiver hardware
operating in the shortwave band and capturing real signals of opportunity.
In summary, the design of the classifier follows a three step approach:
\begin{enumerate}
\item Generation of training data
\item Training of a neural network classifier with backpropagation
\item Evaluation of the trained CNN \textit{on real-world signals}
\end{enumerate}

\subsection{Training Data}

Deep learning is a data-driven approach to classification and thus
requires large amounts of high-quality training data. In this work,
the training data is based on synthetic and high quality real-world
radio signals from available open sources (e.g. \cite{SignalIdentificationGuide.2024,IARUMS.2024}
and others), custom recordings, software generated signals and commercially
available signal libraries. These signals are augmented, i.e. artificially
distorted, to provide diverse and realistic training data. The augmentations
are specifically designed for shortwave signals, such as the Watterson
fading channels, that model ionospheric propagation, or the models
for atmospheric noise. The training data is augmented using the following
signal impairments:
\begin{itemize}
\item Random frequency offset between +/-500 Hz
\item Random phase shift
\item Random sampling rate offset between 0 and 1 \%
\item Bandwidth filter with random excess bandwidth \cite{Scholl.2022}
\item Random SNR between -10 and +25 dB (Gaussian noise)
\item Random introduction of impulsive noise to emulate atmospheric noise
\cite{Scholl.2022}
\item Random channels: 16 Watterson fading models including those defined
in CCIR-520 and ITU 1487 \cite{ITUR.052000,Scholl.2022,C.Watterson.1970}
\end{itemize}
These augmentations enable the neural network to focus on characteristic
signal properties while ignoring typical distortions, like noise,
fading or frequency offsets, that are present in real RF systems \cite{Scholl.2022}.

The dataset consists of complex IQ signals with a sampling rate of
4 kHz, thus covering a bandwidth of approximately 4 kHz, which is
typical for most shortwave signals. The length of each training signal
is 4096 IQ samples, which corresponds to a duration of approximately
1 second. It is assumed, that only one class is present in each training
sample. The complete training dataset contains 7,500 signals per mode,
resulting in a total amount of 1.2 million training samples. The training
dataset covers the 160 HF signal classes listed in Table \ref{tab:Table-of-modes}.
Some exemplary training signals are shown in Figure \ref{fig:Some-exemplary-training}.

\begin{figure}
\begin{centering}
\includegraphics[width=1\columnwidth]{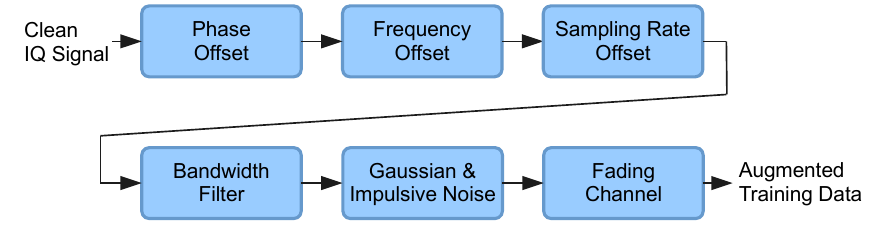}
\par\end{centering}
\begin{centering}
\caption{Training data augmentation}
\par\end{centering}
\end{figure}

\subsection{Neural Network and Training}

CNNs have been successfully applied to various classification problems
for digital signals and are able to provide high accuracy for many
recognition tasks. In addition, they can be efficiently trained using
GPUs and have high representational power \cite{LeCun.1995}. Thus,
they are well suited to a wide range of RF applications, including
signal classification.

The applied neural network follows the typical CNN structure and consists
of an arrangement of 28 layers including convolutional, pooling and
fully connected layers \cite{Geron.October2022}. The convolutional
layers implement non-linear filters, that successively extract and
amplify characteristic features of the input signals. Pooling layers
reduce the length of the signal and act as a decimation-like operation
to force the network to learn more expressive and global features.
The layers use ReLU activation functions and dropout to prevent overfitting.
The input to the network is IQ data, where the I and Q components
are fed into the network as two-channel data. In total, the network
has 1.7 million parameters.

The CNN has been trained for 50 epochs using Adam optimization with
learning rate scheduling. For training and validation, the dataset
is split into two distinct parts: 90~\% for training and 10~\% for
validating the training process.

\subsection{Real-World Test Data}

For deployment and real-world operation, it is important to test the
trained neural network on real-world data. For this purpose, an additional
large test dataset has been collected, that covers different real-world
scenarios. The test data consists exclusively of additional actual
recordings of real signals of opportunity from different SDR receivers
at worldwide locations, such as the Twente WebSDR \cite{PieterTjerkdeBoer.UniversityofTwente},
the KiwiSDR \cite{kiwisdr_network} network and others. The recordings
exhibit varying daytime, season and operating frequency. It further
includes different SNRs and fading channel conditions as well as varying
background noise from the environment in the form of man-made and
atmospheric noise. These variations result in a highly diverse set
of test data, that allows for a meaningful measurement of accuracy
in practical operation. There are no augmentations applied to the
real-world recordings and none of the recordings have been used for
training or training validation. For 143 of the 160 supported modes,
a significant amount of real-world data could be obtained. An overview
of the test data properties is given in Table \ref{fig:Real-World-Test-Data}.

\begin{table}
\begin{centering}
\begin{tabular}{>{\raggedright}m{2.5cm}>{\raggedright}p{5cm}}
\toprule 
{\footnotesize{}Tested Modes} & {\footnotesize{}143 (out of 160)}\tabularnewline
{\footnotesize{}Receiver Hardware} & {\footnotesize{}Kiwi SDR, Airspy HF+, SDR Play, Twente WebSDR, Elad
FDM-S3}\tabularnewline
{\footnotesize{}Locations} & {\footnotesize{}Worldwide}\tabularnewline
{\footnotesize{}Frequencies} & {\footnotesize{}3 - 30 MHz}\tabularnewline
{\footnotesize{}Daytime and Season} & {\footnotesize{}All seasons and varying daytime}\tabularnewline
{\footnotesize{}Signal SNR} & {\footnotesize{}-10 to 25 dB}\tabularnewline
{\footnotesize{}Recording Duration} & {\footnotesize{}>35 hours total}\tabularnewline
\bottomrule
\end{tabular}
\par\end{centering}
\medskip{}

\caption{\label{fig:Real-World-Test-Data}Real-World Test Data}
\end{table}

\section{Classification Results}

\begin{figure}
\begin{centering}
\includegraphics[scale=0.7]{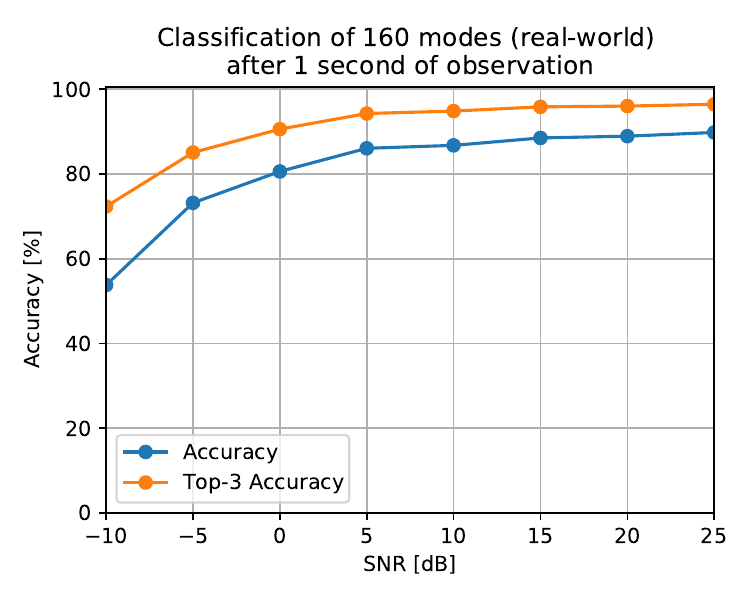}
\par\end{centering}
\caption{\label{fig:results-accuracy}Real-world accuracy and top-3 accuracy
as average over all signal classes}

\end{figure}

In order to measure how well the trained neural network performs in
practice, it is tested on the real-world data set after the training.
Figure \ref{fig:results-accuracy} shows the accuracy and top-3 accuracy
averaged over all modes. The achieved accuracy is around 90~\% for
high SNR values. This means that in 9 of 10 cases the classifier selects
the correct mode out of 160 possible classes, based on only one second
of observation. The top-3 accuracy is approximately 95~\%. In addition,
the classifier is robust to noise and achieves good accuracy even
for signals with lower SNR. Note, that SNR here refers to the full
system bandwidth of 4 kHz.

\begin{figure}
\begin{centering}
\includegraphics[scale=0.6]{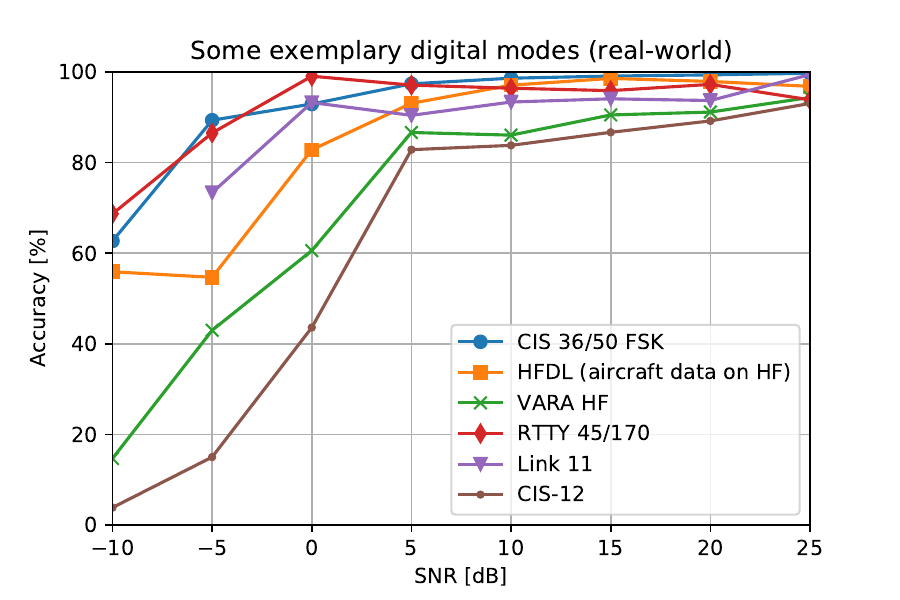}
\par\end{centering}
\caption{\label{fig:results-digital}Real-world accuracy for some exemplary
digital modes}

\end{figure}

\begin{figure}
\begin{centering}
\includegraphics[scale=0.6]{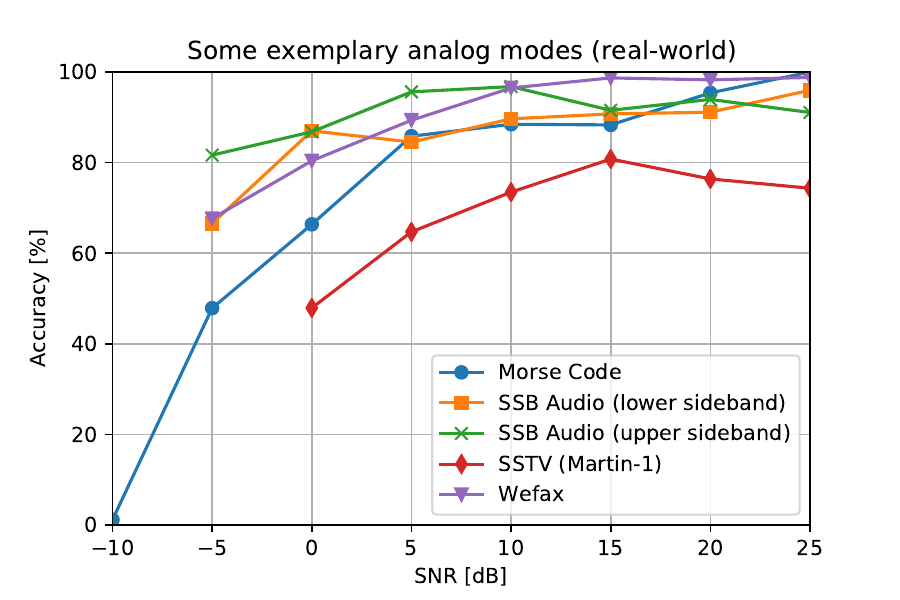}
\par\end{centering}
\caption{\label{fig:results-analog}Real-world accuracy for some exemplary
analog modes}

\end{figure}

\begin{figure}
\begin{centering}
\includegraphics[scale=0.6]{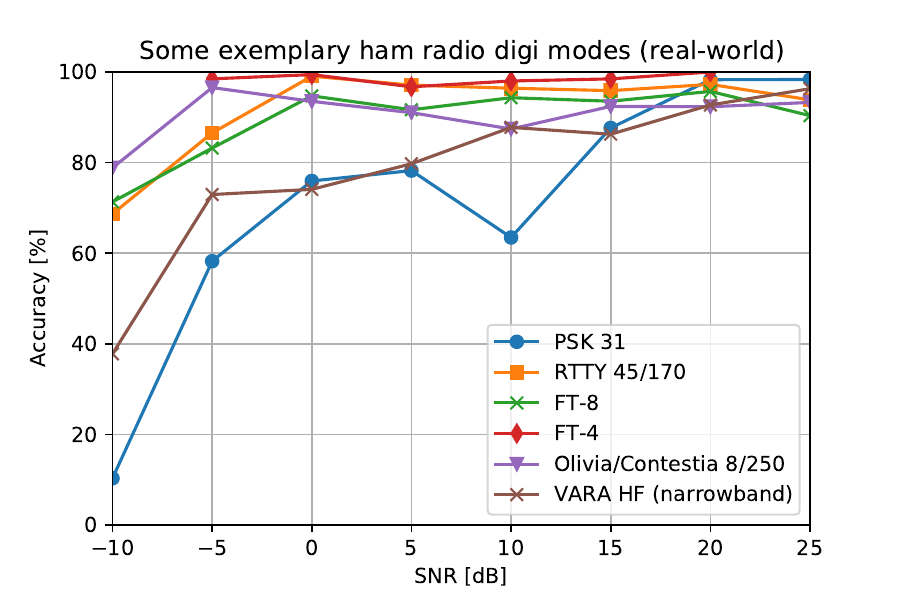}
\par\end{centering}
\caption{\label{fig:results-amateur}Real-world accuracy for some exemplary
amateur radio modes}

\end{figure}

Figures \ref{fig:results-digital}, \ref{fig:results-analog} and
\ref{fig:results-amateur} show more detailed results for a selection
of some common shortwave signal types. Here, some variance over the
different class types can be observed: While some modes achieve accuracies
below 80~\%, others are identified with almost 100~\% accuracy (at
sufficiently high SNR values).

\begin{figure*}
\begin{centering}
\includegraphics[scale=0.75]{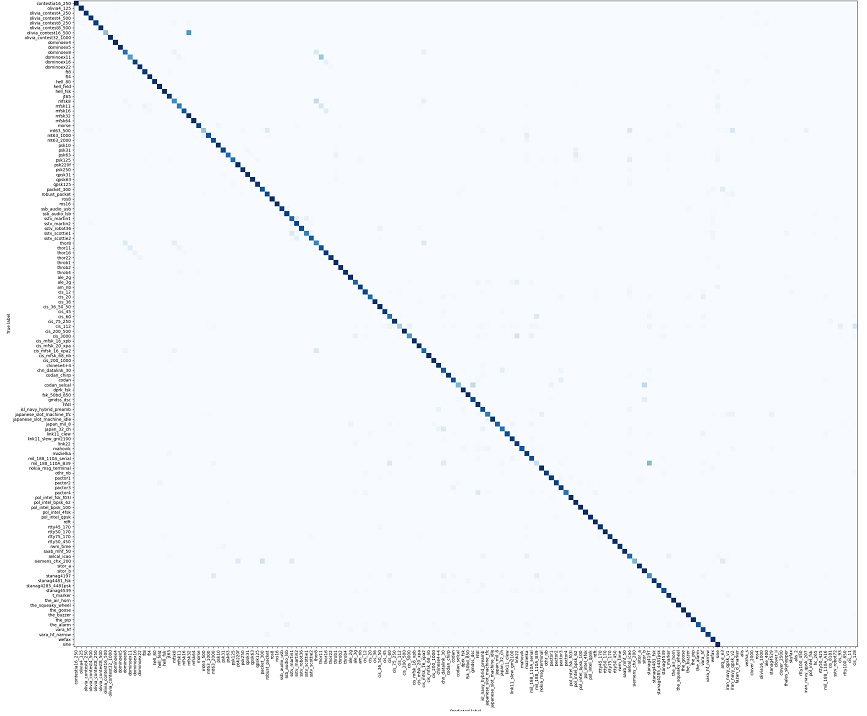}
\par\end{centering}
\caption{\label{fig:results-confusion-1}Confusion matrix for all tested modes
(left). For some of the 160 modes, no test data was available (empty
columns on the right)}
\end{figure*}

\begin{figure*}
\begin{centering}
\includegraphics[scale=0.8]{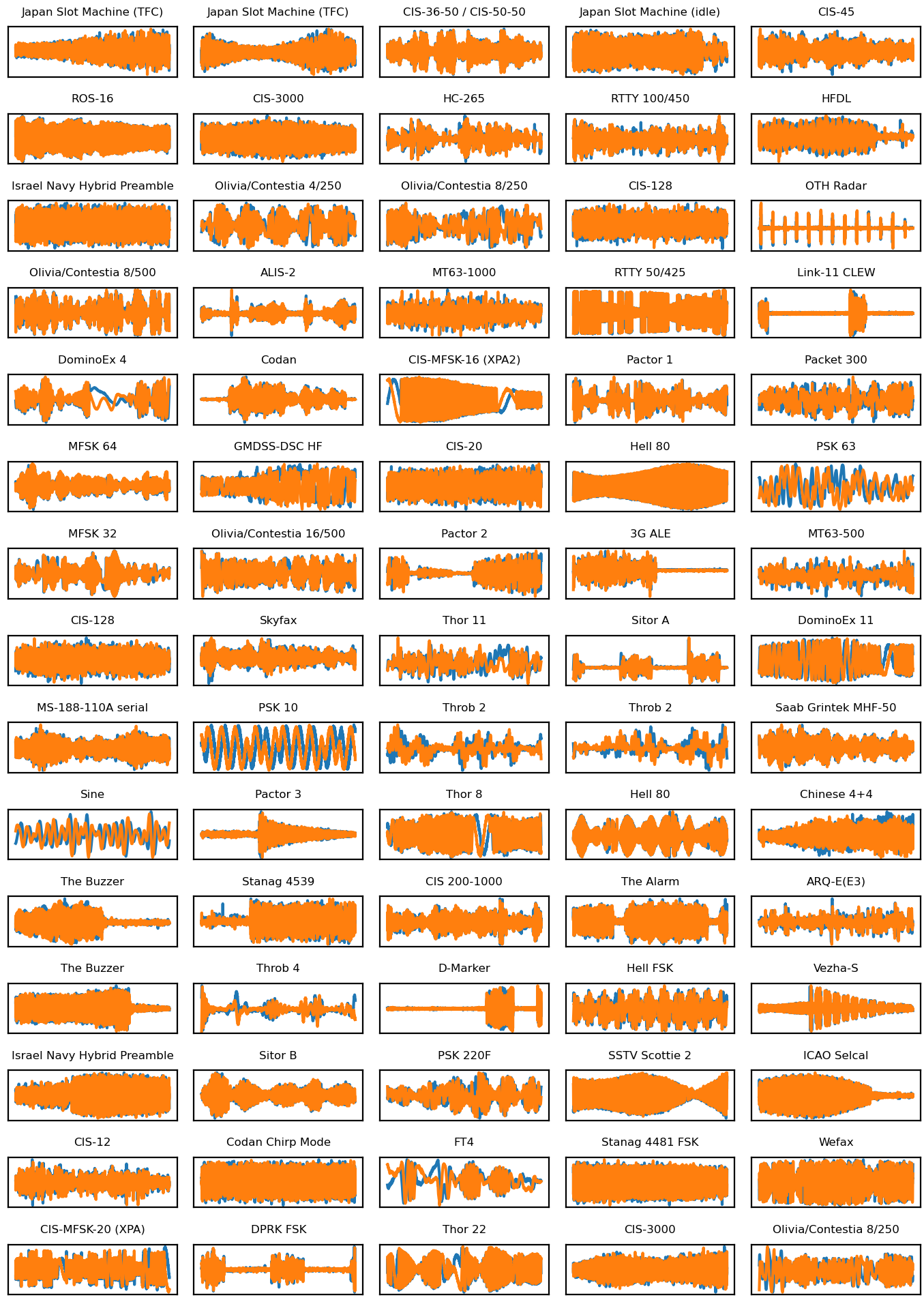}
\par\end{centering}
\caption{\label{fig:Some-exemplary-training}Some exemplary training data samples
(at high SNR for better visualization)}

\end{figure*}

The confusion matrix provides a more detailed picture of the achieved
classification results on real-world data and is shown in Figure \ref{fig:results-confusion-1}.

Although the neural network in general performs well on practical
signals, for some modes the classification accuracy does not approach
100~\% even under high SNR conditions. There are several possible
reasons for this, e.g. the comparably short observation time or the
high similarity of some classes in the time domain, that can lead
to confusions. In addition, the complex structure of deep neural networks
sometimes prevents a clear explanation of incorrect decisions. A number
of techniques, summarized under the term explainable AI are under
investigation to address this shortcoming. 

\section{Summary}

The paper presented an approach to large-scale RF signal classification
with 160 classes using a deep neural network. The work takes into
account the challenges of shortwave band observations, such as typical
channel conditions and the large number of signal types in a loosely
organized band. The neural network has been trained on a large amount
of training data without any manual feature design. The results demonstrate,
that the presented approach can achieve remarkably good accuracy even
for a high number of signal classes and when tested against real-world
signals.

\bibliographystyle{IEEEtran}
\bibliography{literatur_shortwave}

\begin{table*}[t]
\begin{centering}
\begin{tabular}{llll}
\toprule 
2G ALE & DominoEx 8 & Olivia 4/125 & RWM Time\tabularnewline
\midrule
3G ALE & DPRK FSK & Olivia/Contestia 16/500 & Saab Grintek MHF-50\tabularnewline
\midrule
ALE-400 & D-Marker & Olivia/Contestia 32/1000 & ICAO Selcal\tabularnewline
\midrule
ALIS & FSK 50/850 & Olivia/Contestia 4/250 & Siemens CHX-200\tabularnewline
\midrule
ALIS-2 & FT4 & Olivia/Contestia 4/500 & Sine\tabularnewline
\midrule
AM signal & FT8 & Olivia/Contestia 8/250 & Sitor A\tabularnewline
\midrule
ARQ-E(E3) & GMDSS-DSC HF & Olivia/Contestia 8/500 & Sitor B\tabularnewline
\midrule
Chinese 4+4 & HC-265 & OTH Radar & Skyfax\tabularnewline
\midrule
Chinese MIL Datalink 30 & Hell 80 & Packet 300 & Single-Sideband Audio (LSB)\tabularnewline
\midrule
CIS-11 & Hell Feld & Pactor 1 & Single-Sideband Audio (USB)\tabularnewline
\midrule
CIS-112 & Hell FSK & Pactor 2 & SSTV Martin 1\tabularnewline
\midrule
CIS-12 & HFDL & Pactor 3 & SSTV Martin 2\tabularnewline
\midrule
CIS-128 & Iran Navy PSK modem & Pactor 4 & SSTV Robot 36\tabularnewline
\midrule
CIS-20 & Iran Navy PSK modem v1 & Pol Intel 4-FSK & SSTV Robot 72\tabularnewline
\midrule
CIS 200-1000 & Iran Navy PSK modem v2 & Pol Intel BPSK P03k & SSTV Scottie 1\tabularnewline
\midrule
CIS-200-500 & Israel Navy Hybrid Preamble & Pol Intel BPSK P03i & SSTV Scottie 2\tabularnewline
\midrule
CIS-3000 & Japan 32-Channel & Pol Intel FSK & Stanag 4197\tabularnewline
\midrule
CIS-36 & Japan 8-Channel & Pol Intel FSK F03l & Stanag 4285\tabularnewline
\midrule
CIS-36-50 / CIS-50-50 & Japan Slot Machine (idle) & Pol Intel QPSK & Stanag 4481 FSK\tabularnewline
\midrule
CIS-45 & Japan Slot Machine (TFC) & PSK 10 & Stanag 4529\tabularnewline
\midrule
CIS-60 & JT65 & PSK 125 & Stanag 4539\tabularnewline
\midrule
CIS-75-250 & Link-11 CLEW & PSK 220F & T-Marker\tabularnewline
\midrule
CIS-8181 & Link-11 SLEW / GM2100 & PSK 250 & Thales Skyhopper\tabularnewline
\midrule
CIS-93 & Link-22 & PSK 31 & The Air Horn\tabularnewline
\midrule
CIS-MFSK-16 (XPA2) & Mahovik & PSK 63 & The Alarm\tabularnewline
\midrule
CIS-MFSK-16 (XPB) & Mazielka & QPSK 125 & The Buzzer\tabularnewline
\midrule
CIS-MFSK-20 (XPA) & MFSK 11 & QPSK 31 & The Goose\tabularnewline
\midrule
CIS MFSK-68 & MFSK 16 & QPSK 63 & The Pip\tabularnewline
\midrule
Clover 2000 & MFSK 32 & RDFT & The Squeaky Wheel\tabularnewline
\midrule
Clover 2500 & MFSK 64 & Robust Packet & Thor 11\tabularnewline
\midrule
Clover II & MFSK 8 & ROS-16 & Thor 16\tabularnewline
\midrule
Codan & MS-188-110A A16 & ROS-4 & Thor 22\tabularnewline
\midrule
Codan Chirp Mode & MS-188-110A B39 & ROS-8 & Thor 8\tabularnewline
\midrule
Codan Selcall & MS-188-110A serial & RTTY 100/450 & Throb 1\tabularnewline
\midrule
Contestia 16/250 & Morse Code & RTTY 100/850 & Throb 2\tabularnewline
\midrule
DominoEx 11 & MT63-1000 & RTTY 45/170 & Throb 4\tabularnewline
\midrule
DominoEx 16 & MT63-2000 & RTTY 50/170 & Vara HF Std\tabularnewline
\midrule
DominoEx 22 & MT63-500 & RTTY 50/425 & Vara HF Narrow\tabularnewline
\midrule
DominoEx 4 & Nokia Adaptive MSG Terminal & RTTY 50/450 & Vezha-S\tabularnewline
\midrule
DominoEx 5 & Olivia 16/1000 & RTTY 75/170 & Wefax\tabularnewline
\bottomrule
\end{tabular}
\par\end{centering}
\medskip{}

\caption{\label{tab:Table-of-modes}Table of supported modes}
\end{table*}

\end{document}